\documentclass[conference]{IEEEtran}
\IEEEoverridecommandlockouts
\pdfoutput=1
\usepackage{cite}
\usepackage{amsmath,amssymb,amsfonts}
\usepackage{algorithmic}
\usepackage{graphicx}
\usepackage{hyperref}
\usepackage{subcaption}
\usepackage{textcomp}
\usepackage{xcolor}
\usepackage{siunitx}
\usepackage{comment}
\usepackage{xspace}
\usepackage{url}

\renewcommand{\theenumi}{\roman{enumi}}

\usepackage{booktabs}

\newcommand{\myparagraph}[1]{\vspace{1mm}\noindent\textbf{#1}:}

\def\BibTeX{{\rm B\kern-.05em{\sc i\kern-.025em b}\kern-.08em
    T\kern-.1667em\lower.7ex\hbox{E}\kern-.125emX}}
    
\makeatletter
\newcommand{\linebreakand}{%
  \end{@IEEEauthorhalign}
  \hfill\mbox{}\par
  \mbox{}\hfill\begin{@IEEEauthorhalign}
}
\makeatother
    
\begin{document}

\title{The Hitachi-JHU DIHARD III System: Competitive End-to-End Neural Diarization and X-Vector Clustering Systems Combined by DOVER-Lap}

\author{
    \IEEEauthorblockN{Shota Horiguchi$^1$\quad Nelson Yalta$^1$\quad Paola Garc\'{i}a$^2$\quad Yuki Takashima$^1$\quad Yawen Xue$^1$\\Desh Raj$^2$\quad Zili Huang$^2$\quad Yusuke Fujita$^1$\quad Shinji Watanabe$^2$\quad Sanjeev Khudanpur$^2$}
    \IEEEauthorblockA{
        \textit{$^1$\textit{Hitachi, Ltd. Research \& Development Group}, Tokyo, Japan} \\
        \textit{$^2$Center for Language and Speech Processing, Johns Hopkins University, Baltimore, MD, USA}\\
        \{shota.horiguchi.wk, nelson.yalta.wm\}@hitachi.com, lgarci27@jhu.edu
    }
}

\maketitle

\begin{abstract}
This paper provides a detailed description of the Hitachi-JHU system that was submitted to the Third DIHARD Speech Diarization Challenge.
The system outputs the ensemble results of the five subsystems: two x-vector-based subsystems, two end-to-end neural diarization-based subsystems, and one hybrid subsystem.
We refine each system and all five subsystems become competitive and complementary.
After the DOVER-Lap based system combination, it achieved diarization error rates of 11.58\% and 14.09\% in Track 1 full and core, and 16.94\% and 20.01\% in Track 2 full and core, respectively.
With their results, we won second place in all the tasks of the challenge.
\end{abstract}


\section{Notable highlights}
This technical report describes the Hitachi-JHU system submitted to the Third DIHARD Speech Diarization Challenge \cite{ryant2020third}.
We mainly focused our efforts on how we can pick the best of diarization based on x-vector clustering and end-to-end neural speaker diarization (EEND).
The highlights of our systems are as follows:
\begin{itemize}
    \item Two x-vector-based systems incorporated with VBx clustering and heuristic overlap assignment. One is based on a time-delay neural network (TDNN) based x-vector extractor following the winning system of the DIHARD II Challenge \cite{landini2019but,landini2020but}. The other is based on Res2Net-based x-vector extractors, which won the VoxCeleb Speaker Recognition Challenge 2020 \cite{xiao2020microsoft}.
    \item Two EEND-based subsystems, each of which is the extension of the original self-attentive EEND \cite{fujita2019end2} to output diarization results of a variable number of speakers, with improved inference.
    \item A hybrid subsystem of x-vector clustering and EEND, in which update the results of x-vector clustering using EEND as post-processing \cite{horiguchi2021endtoend}.
    \item Modified DOVER-Lap \cite{raj2021doverlap} to combine the results from five subsystems above.
    \item Self-supervised adaptation of the EEND model.
\end{itemize}

\section{Data resources}
\autoref{tbl:corpora} summarizes the corpora we used to train our models which compose our diarization system.
We briefly explain each corpus below.

\begin{itemize}
    \item DIHARD III: focused on ``hard'' speaker diarization, contains 5-10 minute utterances selected from 11 conversational domains, each including approximately 2 hours of audio~\cite{ryant2020third}.
    \item VoxCeleb 1: a large-scale speaker identification dataset with 1,251 speakers and over 100,000 utterances, collected ``in the wild''~\cite{nagrani2020voxceleb}.
    \item VoxCeleb 2: a speaker recognition dataset that contains over a million utterances from over 6,000 speakers under noisy and unconstrained conditions~\cite{nagrani2020voxceleb}.
    \item Switchboard-2 (Phase I, II, III), Switchboard Cellular (Part 1, 2): English telephone conversation datasets. Their LDC catalog numbers are LDC98S75, LDC99S79, LDC2002S06, LDC2001S13, and LDC2004S07, respectively.
    \item NIST Speaker Recognition Evaluation (2004, 2005, 2006, 2008): also telephone conversations but not limited to English, which are composed of the following LDC corpora: LDC2006S44, LDC2011S01, LDC2011S04, LDC2011S09, LDC2011S10, LDC2012S01, LDC2011S05, LDC2011S08.
    \item MUSAN: publicly available corpus that consists of music, speech, and noise \cite{snyder2015musan}. The music and noise portions are sometimes used for data augmentation.
\end{itemize}

\begin{table*}[t]
    \centering
    \caption{Corpora we used to train the models in our system.}
    \label{tbl:corpora}
    \resizebox{\linewidth}{!}{%
    \begin{tabular}{@{}lcccccccc@{}}
        \toprule
        &\multicolumn{2}{c}{VAD}&\multicolumn{2}{c}{X-vector extractor}\\\cmidrule(lr){2-3}\cmidrule(lr){4-5}
        Corpus & SincNet&TDNN & TDNN & Res2Net & PLDA & Overlap detector & EEND-EDA & SC-EEND\\\midrule
        DIHARD III development set \cite{ryant2020third}&\checkmark& \checkmark & && \checkmark & \checkmark & \checkmark & \checkmark\\
        DIHARD III evaluation set \cite{ryant2020third} (with pseudo labels)&&&&&&&\checkmark\\
        VoxCeleb 1 \cite{nagrani2020voxceleb}&& &\checkmark&\checkmark\\
        VoxCeleb 2 \cite{nagrani2020voxceleb}&& &\checkmark&\checkmark\\
        Switchboard-2 Phase I, I\hspace{-.1em}I, I\hspace{-.1em}I\hspace{-.1em}I&& &&&&&\checkmark&\checkmark\\
        Switchboard Cellular Part 1, 2&& &&&&&\checkmark&\checkmark\\
        NIST Speaker Recognition Evaluation 2004, 2005, 2006, 2008&& &&&&&\checkmark&\checkmark\\
        MUSAN corpus&&\checkmark &\checkmark&\checkmark&&&\checkmark&\checkmark\\
        \bottomrule
    \end{tabular}%
    }
\end{table*}

\section{Detailed description of algorithm}

\subsection{Voice Activity Detector}
We employed two voice activity detectors (VAD): SincNet-based VAD \cite{Lavechin2020} and TDNN-based VAD.

\myparagraph{SincNet-based VAD}
Our SincNet-based VAD is implemented using the pyannote \cite{Bredin2020} framework. This VAD model learns to detect speech from the raw speech using a combination of a SincNet \cite{ravanelli2018speaker} followed by BiLSTM layers and fully connected layers. For our experiments, we employed the default configuration provided by pyannote: a SincNet with 80 channels and 251 dims of kernel size, two BiLSTM layers with 128 cell dimensions, and two fully connected layers of 128 dimensions. We trained the model using the DIHARD III development set for 300 epochs.

\myparagraph{TDNN-based VAD}
Our TDNN VAD is based on the example from Kaldi \cite{povey2011kaldi} recipe. The acoustic feature we use is $40$-dim MFCC, and the left and right 2 frames are appended to generate the 200-dim input features. The model first transforms the input features with linear discriminant analysis (LDA) estimated with the VAD labels. Then the transformed features pass through five TDNN blocks. Each TDNN block consists of a TDNN layer, a Rectified Linear Unit (ReLU), and a batch normalization layer. In the last two TDNN blocks, to capture long temporal contexts, the mean vector for neighboring frames is computed as an additional input. Finally, a linear layer is used to predict the probability for each frame.
The model was trained on the DIHARD III development set for around 10 epochs. We augmented the training data with the noise, music, and babble from the MUSAN\cite{snyder2015musan} corpus and created some reverberated speech with simulated room impulse responses \cite{ko2017study}.

The final results of VAD were calculated by averaging posterior probabilities from the two models, followed by thresholding and median filtering.
As shown in the \autoref{tbl:vad}, posterior averaging of the two systems achieved the best trade-off between false alarms and missed speech than the individuals.

\begin{table}[t]
    \centering
    \caption{VAD performance on the DIHARD III development set.}
    \label{tbl:vad}
    \begin{tabular}{@{}lcc@{}}
        \toprule
        Method&False alarm (\%)&Missed speech (\%)\\\midrule
        SincNet-based VAD & 2.78 & 2.51\\
        TDNN-based VAD & 2.85 & 2.80\\
        Posterior average&2.58&2.55\\
        \bottomrule
    \end{tabular}
\end{table}

\subsection{X-vector-based subsystems}

\subsubsection{TDNN  (\textbf{System (1)})}\label{sec:tdnn}
The TDNN x-vector-based system consists of two main parts: TDNN extractor and the VBx clustering.

\myparagraph{TDNN-based extractor}
It employs 40-dimensional filterbanks, with a \SI{25}{\ms} window and \SI{15}{\ms} frame shift. 
These features are used for the embedding extraction as in~\cite{diez2018speaker}.
The x-vector was trained using a TDNN with a \SI{1.5}{\second} window with frame shift of \SI{0.25}{\second}.
The TDNN extractor consists of four TDNN-ReLU layers each of them followed by a dense-ReLU.
Then, two dense-ReLU layers are incorporated before a pooling layer; a final dense-ReLU is included from which 512-dimension embeddings are computed. A dense-softmax concludes this TDNN architecture \cite{snyder2019speaker}.

\myparagraph{VBx clustering}
To eliminate the need for a tuned agglomerative hierarchical clustering (AHC) stopping threshold, we perform VBx-clustering after AHC \cite{diez2018speaker}.
The VBx-clustering is a simplified variational Bayes diarization. It follows a hidden Markov model (HMM), in which the state represents a speaker, and the state transitions correspond to speaker turns.
The state distributions, or emission probabilities, are Gaussian mixture models constrained by eigenvoice matrix.
Each speaker has a probability of $P_\text{loop}$ when the HMM ends up back in the same state. The initialization for this system is a probabilistic LDA (PLDA) model. For our experiments, this PLDA is the result of the interpolation of the VoxCeleb PLDA and the in-domain DIHARD III PLDA. Both PLDAs were centered and whitened using DIHARD III development set. 

For the TDNN-based system, the x-vectors were projected from 512 dim to 220 using an LDA, the PLDA interpolation regulated by an alpha was set to 0.50, and the value for $P_\text{loop}$ to 0.80.

We finally applied overlap assignment, which is described in \autoref{sec:overlap_assignment}, to obtain the final diarization results from this subsystem.

\subsubsection{Res2Net (\textbf{System (2)})}
Initially proposed for image recognition, Res2Net was applied to speaker embedding because it provides highly accurate speaker clustering \cite{zhou2021resnext}. 

The Res2Net-based extractor uses the default configuration described in \cite{zhou2021resnext}. The Res2Net uses 80 log-filterbank dimensions as input, a multi head-attentive pooling with attention heads set to 16 that learns to weight each frame, and additive angular margin Softmax (AAM) \cite{Deng2019AAM} with margin of 0.1 and scale of 30 as a training criterion. For our experiments, we employed four extractors:

\begin{enumerate}
    \item \textbf{Res2Net-UN}: This extractor employs the default configuration of a Res2Net with 23 layers, utterance normalization, $\log_{10}$ compression, AAM margin of 0.1, and AAM scale of 30.
    \item \textbf{Res2Net-BN}: This extractor is similar to Res2Net-UN, with a batch normalization layer instead of utterance normalization and $\ln$ compression.
    \item \textbf{Res2Net-BN-Large}: This extractor uses a Res2Net with 50 layers with a similar configuration as Res2Net-BN.
    \item \textbf{Res2Net-UN-Large}: This extractor uses a Res2Net with 50 layers and a similar configuration as Res2Net-UN. Additionally, it uses SpecAugment \cite{park2019specaugment} for data augmentation.
\end{enumerate}

We employed the VoxCeleb 1 and VoxCeleb 2 sets \cite{nagrani2020voxceleb} as training that provided 7323 speakers and over 1M of recordings. We augmented the data following similar data augmentation as the Kaldi recipe for VoxCeleb\footnote{https://github.com/kaldi-asr/kaldi/blob/master/egs/voxceleb/v2}. Each audio recording is randomly chunked into subsegments of length between \SI{2.0}{\second} and \SI{4.5}{\second} that are feed into the models.

Similarly to the TDNN-based system, the 128-dimension embeddings, were passed through and LDA without reduction and used a PLDA interpolation regulated by an alpha was set to 0.10, and the value for $P_\text{loop}$ to 0.80.

Once the results from the four extractors were obtained, we combined the results using modified DOVER-Lap, which is explained in \autoref{sec:system_fusion}.

\subsubsection{Overlap detection and assignment}\label{sec:overlap_assignment}
For the Res2Net and the TDNN x-vector subsystems, we used a similar approach to perform overlap detection like the one shown for the SincNet-based VAD, with the only difference that the classifier will distinguish between overlapping speech versus non-overlapping speech.
We assigned the closest other speaker in the time axis as the second speaker for each detected frame.

\autoref{tbl:x-vector} shows the diarization error rates (DERs) and Jaccard error rates (JERs) on the DIHARD III development set using the x-vector-based subsystems.

\begin{table*}[t]
    \centering
    \caption{DERs / JERs (\%) of x-vector-based subsystems on the DIHARD III development set.}
    \label{tbl:x-vector}
    \begin{minipage}[t]{0.30\linewidth}
        \centering
        \subcaption{TDNN (System (1))}
        \scalebox{0.97}{%
        \begin{tabular}{@{}lc@{}}
            \toprule
            Method & DER / JER (\%)\\\midrule
            x-vector + VBx & 16.33 / 34.18 \\
            x-vector + VBx + OvlAssign & 13.87 / 32.73\\
            \bottomrule
        \end{tabular}%
        }
    \end{minipage}
    \hfill
    \begin{minipage}[t]{0.68\linewidth}
        \centering
        \subcaption{Res2Net (System (2))}
        \scalebox{0.97}{%
        \begin{tabular}{@{}lcccc@{}}
            \toprule
            &\multicolumn{4}{c}{DER / JER (\%)}\\\cmidrule(l){2-5}
            Method&Res2Net-BN&Res2Net-UN&Res2Net-BN-Large&Res2Net-UN-Large\\\midrule
            x-vector + VBx&17.24 / 37.12&17.04 / 36.17 & 16.85 / 35.86 & 17.08 / 35.95\\
            x-vector + VBx + OvlAssign&14.89 / 35.64 & 14.72 / 34.65 & 14.56 / 34.31 & 14.74 / 34.40\\\midrule
            Modified DOVER-Lap&\multicolumn{4}{c}{14.04 / 34.29}\\
            \bottomrule
        \end{tabular}%
        }
    \end{minipage}
\end{table*}

\subsection{EEND-based subsystems}
We employed EEND-EDA \cite{horiguchi2020endtoend} and SC-EEND \cite{fujita2020neural} as EEND-based subsystems, each of which can handle a flexible number of speakers.
The inputs to the EEND-based models were based on log-Mel filterbanks but with different configurations for each model.
For EEND-EDA, 23-dimensional log-Mel filterbanks was extracted with frame length of \SI{25}{\ms} and frame shift of \SI{10}{\ms} from \SI{8}{\kHz} recordings. Each filterbanks were then concatenated with those from the left and right seven frames to construct 345-dimensional features.
We subsampled them by a factor of 10 to obtain input features for each \SI{100}{\ms} during training and that of five to obtain features for each \SI{50}{\ms} during inference.
For SC-EEND, we used 40-dimensional log-Mel filterbanks from \SI{16}{\kHz} recordings and concatenated the left and right 14 frames to construct 1160-dimensional features.
The subsampling factor was set to 20 during pretraining using simulated mixtures and 10 during adaptation and inference.

\subsubsection{EEND-EDA (\textbf{System (3)})}\label{sec:eend_eda}
EEND-EDA \cite{horiguchi2020endtoend} calculates posteriors by dot products between time-frame-wise embeddings and speaker-wise attractors, which are calculated from the embeddings using encoder-decoder attractor calculation module (EDA).
The training procedure depends on simulated mixtures summarized in \autoref{tbl:sim} and the DIHARD III corpus.
We created them using the script provided in the EEND repository\footnote{\url{https://github.com/hitachi-speech/EEND/blob/master/egs/callhome/v1/run_prepare_shared_eda.sh}} with various $\beta$ values shown in \autoref{tbl:sim}, which determines the average duration of silence between utterances.
We first trained the model using Sim2spk for 100 epochs, then finetuned it on the concatenation of Sim1spk to Sim5spk for another 75 epochs, and finally adapted it on the DIHARD III development set for 200 epochs.
We used Adam optimizer \cite{kingma2015adam} for all the training, but with Noam scheduler \cite{vaswani2017attention} that set the warm-up steps to 100,000 iterations for training on simulated mixtures and with a fixed learning rate of $1\times10^{-5}$ for adaptation.

\begin{table}[t]
    \centering
    \caption{Simulated mixtures used for EEND-EDA training. Sim1spk, Sim2spk, Sim3spk, and Sim4spk are the same as ones used in the EEND-EDA paper \cite{horiguchi2020endtoend}.}
    \label{tbl:sim}
    \begin{tabular}{@{}lcccc@{}}
        \toprule
        Dataset & \#Spk & \#Mixtures & $\beta$ & Overlap ratio (\%)\\\midrule
        Sim1spk & 1 & 100,000 & 2 & 0.0 \\
        Sim2spk & 2 & 100,000 & 2 & 34.1\\
        Sim3spk & 3 & 100,000 & 5 & 34.2\\
        Sim4spk & 4 & 100,000 & 9 & 31.5\\
        Sim5spk & 5 & 100,000 & 13 & 30.3\\
        \bottomrule
    \end{tabular}
\end{table}

During inference, we used the dereverberated audio using weighted prediction error (WPE) \cite{nakatani2010speech}. We estimated a dereverberation filter on Short Time Fourier Transform (STFT) spectrum using the entire audio recording as an input block. The STFT features are computed using a window of \SI{32}{\ms} (512 dims) and shifting of \SI{8}{\ms} (128 dims). Using 5 iterations, we set the prediction delay and the filter length to 3 and 30, respectively, for \SI{16}{\kHz}.

Because EEND-based models conduct speaker diarization and voice activity detection simultaneously, they must be incorporated with oracle speech segments (for Track 1) or accurate external VAD (for Track 2) to fit the DIHARD tasks.
Thus, once the diarization results were obtained using the EEND-EDA model, we filtered false alarms and recovered missed speech by assigning the speakers with the highest posterior probabilities using VAD.
In this paper, we call these procedures VAD post-processing.

Even if the adaptation was based on the DIHARD III development set, which contains mixtures of at most 10 speakers, it is difficult to produce diarization results of more than five speakers because its pretraining was based on mixtures in which include at most five speakers.
Therefore, we produce diarization results for more than five speakers using an iterative inference as follows:
\begin{enumerate}
    \item decide the maximum number of speakers $K(\leq5)$ to decode,
    \item decode at most $K$ speaker's diarization results,
    \item stop inference if the estimated number of speakers is less than $K$ otherwise continue to the next step,
    \item select frames in which all the decoded speakers are inactive and back to i),
\end{enumerate}
We varied $K\in\left\{1,2,3,4,5\right\}$ at the first iteration and fixed it to $5$ from the second iteration.
Finally, the five estimated results are combined using the modified DOVER-Lap described in \autoref{sec:system_fusion} to obtain the final results of the EEND-EDA-based system.

\autoref{tbl:eda} shows DERs and JERs of the EEND-EDA-based and SC-EEND-based subsystems.
It clearly indicates that the VAD post-processing and the iterative inference improved the diarization performance.

\begin{table}[t]
    \centering
    \caption{DERs and JERs (\%) on the DIHARD III development set using EEND-based models. FA: false alarm, MI: missed speech.}
    \label{tbl:eend}
    \begin{minipage}[t]{\linewidth}
        \centering
        \subcaption{EEND-EDA (System (3))}
        \label{tbl:eda}
        \begin{tabular}{@{}lcc@{}}
            \toprule
            Method&DER&JER\\\midrule
            EEND-EDA&18.77&38.98\\
            + filter FA&17.33&37.92\\
            + recover MI&13.08&35.38\\
            + iterative inference ($K=5$)&13.35&34.19\\
            + iterative inference ($K\in\{1,\dots,5\}$) \& DOVER-Lap&\textbf{12.92}&\textbf{33.85}\\
            \bottomrule
        \end{tabular}
    \end{minipage}
    \\\bigskip
    \begin{minipage}[t]{\linewidth}
        \centering
        \subcaption{SC-EEND (System (4))}
        \label{tbl:sceend}
        \begin{tabular}{@{}lcc@{}}
            \toprule
            Method&DER&JER\\\midrule
            SC-EEND&18.61&39.19\\
            + filter FA&16.02&37.46\\
            + recover MI&\textbf{13.13}&\textbf{35.35}\\
            \bottomrule
        \end{tabular}
    \end{minipage}
\end{table}

\subsubsection{SC-EEND (\textbf{System (4)})}
SC-EEND is a model which estimates each speaker's speech activities one-by-one, conditioned on the previously estimated speech activities.
We used stacked Conformer encoders \cite{gulati2020conformer} instead of Transformer encoders that used in the original SC-EEND.
The model was firstly trained on simulated mixtures, each of which contains at most four speakers, for 100 epochs using Adam optimizer with the same scheduler as in EEND-EDA.
Then, the model was initialized with the average weights of the last 10 epochs and trained again on the simulated mixtures for additional 100 epochs.
Finally, the model was adapted on the DIHARD III development set from the average weights of the last 10 epochs of the second-round pretraining for additional 200 epochs using Adam optimizer with the fixed learning rate of $1\times10^{-5}$.
The details of the simulated mixtures are described in the SC-EEND paper \cite{fujita2020neural}.

For SC-EEND, we also used dereverberated audio and applied VAD post-processing (filtering false alarms and recovering missed speech) as described in \autoref{sec:eend_eda}.
However, the Conformer encoders have order dependency so that we cannot conduct the decoding process only for the selected frames that are not always equally spaced along the time axis.
Therefore, we did not apply the iterative inference for the SC-EEND model.
The results of SC-EEND with step-by-step improvement by using VAD post-processing are shown in \autoref{tbl:sceend}.

\subsection{Hybrid subsystem (\textbf{System (5)})}
We also used EEND as post-processing (EENDasP) \cite{horiguchi2021endtoend} to refine diarization results obtained from the TDNN-based x-vectors described in \autoref{sec:tdnn}.
In EENDasP, two speakers from the results are iteratively selected and their results are updated using the EEND model.
In the original paper, the EEND-EDA model was trained to output only two-speaker results, but we used the first two speakers' output from the model trained in \autoref{sec:eend_eda} for our system.
By applying EENDasP for TDNN-based x-vectors with VBx clustering but without heuristic overlap assignment, DER was improved from \SI{16.33}{\percent} to \SI{12.63}{\percent}.

\subsection{System fusion}\label{sec:system_fusion}
To combine multiple diarization results, we used DOVER-Lap \cite{raj2021doverlap} with a modification.
The original DOVER-Lap assigns uniformly-divided regions for each speaker if the multiple speakers are weighted equally in the label voting stage.
However, we found that it leads to an increase in missed speech.
This is obvious by considering the case when the same three hypotheses with overlaps are input to DOVER-Lap.
The speakers included in the hypotheses are always tied in this case; thus, overlapped regions in the hypotheses are divided to be assigned for each speaker, which results in the combined hypothesis with no overlap.
Thus, we assigned all the tied speakers to the regions without any division.

When we combine diarization results from various systems, we sometimes know that some systems are highly accurate and others are not so.
Therefore, we introduced hypothesis-wise manual weighting to DOVER-Lap.
The original DOVER \cite{stolke2019dover} and DOVER-Lap, the input hypotheses are ranked by their average DER to all the other hypotheses.
In other words, the hypotheses $H_1,\dots,H_k,\dots,H_K$ are ranked by following score $s_k$:
\begin{align}
    s_k=\frac{1}{K-1}\sum_{k'\in\left\{1,\dots,K\right\}, k\neq k'}\mathit{DER}\left(H_k,H_{k'}\right),
    \label{eq:ranking}
\end{align}
where $\mathit{DER}(H_k,H_{k'})$ is the function to calculate diarization error rate from the reference $H_k$ and estimation $H_{k'}$.
In our system, we used $w_k s_k$ instead of $s_k$, where $w_k\in\mathbb{R}_{+}$ is a weighing value, to control the importance of each hypothesis.

\autoref{tbl:doverlap} shows DERs and breakdown on the DIHARD III development set.
Note that the manual weighting was only used to combine five hypotheses in System (9) and not used for combine the Res2Net-results in System (1), EEND-EDA iterative inference in Systems (3) and (7), and the five-system fusion for System (6) due to time constraints.
The weights to combine Systems (1)(2)(4)(7)(9) were set to $w_{(1)}=2$, $w_{(2)}=2$, $w_{(4)}=1$, $w_{(7)}=4$, $w_{(9)}=3$, which were determined by using the development set.

\begin{table}[t]
    \centering
    \caption{Comparison between the original and modified DOVER-Lap on the DIHARD III development set. MI: missed speech, FA: false alarm, CF: speaker confusion.}
    \label{tbl:doverlap}
    \resizebox{\linewidth}{!}{%
    \begin{tabular}{@{}lcccc@{}}
        \toprule
        Method&MI&FA&CF&DER\\\midrule
        (1) TDNN-based x-vector + VBx + OvlAssign&5.36&1.93&6.58&13.87\\
        (2) Res2Net-based x-vector + VBx + OvlAssign&5.47&1.89&6.68&14.04\\
        (3) EEND-EDA &6.54&1.36&5.02&12.92\\
        (4) SC-EEND &4.85&1.96&6.32&13.13\\
        (5) TDNN-based x-vector + VBx + EENDasP&6.53&1.32&4.79&12.63\\\midrule
        DOVER-Lap &6.96&0.77&4.33&12.07\\
        Modified DOVER-Lap (System (6)) &5.53&0.93&4.27&10.73\\
        Modified DOVER-Lap + manual weighting&5.54&0.93&4.21&10.68\\
        \bottomrule
    \end{tabular}%
    }
\end{table}

\subsection{Self-supervised adaptation}\label{sec:self_supervised_adaptation}
After the first system fusion, we applied self-supervised adaptation (SSA) for the EEND-EDA model.
The estimated results were used as the pseudo labels for the DIHARD III evaluation set,
We redid the adaptation step in \autoref{sec:eend_eda} on the concatenation of the DIHARD III development set with the ground truth labels and the evaluation set with the pseudo labels.
With the new model, we placed the results of the EEND-EDA (System (3)), EENDasP (System (5)), and DOVER-Lap (System (6)).
Note that we used different pseudo labels for Track 1 and Track 2 because the oracle VAD was only available on Track 1.

\section{Results}

\begin{table*}[t]
    \centering
    \caption{DERs / JERs (\%) on Track 1 \& 2.}
    \label{tbl:result}
    \resizebox{\linewidth}{!}{%
    \begin{tabular}{@{}lcccccccc@{}}
        \toprule
        &\multicolumn{4}{c}{Track 1 (w/ oracle VAD)}&\multicolumn{4}{c}{Track 2 (w/o oracle VAD)}\\\cmidrule(lr){2-5}\cmidrule(l){6-9}
        &\multicolumn{2}{c}{Dev}&\multicolumn{2}{c}{Eval}&\multicolumn{2}{c}{Dev}&\multicolumn{2}{c}{Eval}\\\cmidrule(lr){2-3}\cmidrule(lr){4-5}\cmidrule(lr){6-7}\cmidrule(l){8-9}
        System & full & core & full & core & full & core & full & core \\\midrule
        Baseline \cite{ryant2020third}&19.41 / 41.66&20.25 / 46.02&19.25 / 42.45& 20.65 / 47.74&21.71 / 43.66&22.28 / 47.75&25.36 / 46.95&27.34 / 51.91\\\midrule
        (1) TDNN-based x-vector + VBx + OvlAssign&13.87 / 32.73&14.88 / 36.72&15.65 / 33.71&18.20 / 38.42&17.61 / 36.03&18.64 / 39.92&21.47 / 37.83&24.58 / 42.02\\
        (2) Res2Net-based x-vector + VBx + OvlAssign&14.04 / 34.29&15.18 / 38.80&15.81 / 35.53&18.47 / 40.47&17.26 / 37.17&18.39 / 41.56&21.37 / 39.59 & 24.64 / 44.49\\
        (3) EEND-EDA&12.92 / 33.85&13.95 / 35.37&13.95 / 35.37&17.28 / 41.97&15.90 / 35.94&18.50 / 41.71&19.04 / 38.89&22.84 / 45.27\\
        (4) SC-EEND&13.13 / 35.35&16.05 / 41.80&15.16 / 38.62&19.14 / 46.04&16.16 / 37.52&19.00 / 43.74&20.30 / 42.19&24.75 / 49.36\\
        (5) TDNN-based x-vector + VBx + EENDasP &12.63 / 31.52&14.61 / 36.28&13.30 / 33.02&15.92 / 38.29&15.94 / 34.11&18.09 / 38.97&18.13 / 35.82&21.31 / 40.78\\
        (6) DOVER-Lap of (1)(2)(3)(4)(5)&10.73 / 31.39&12.56 / 36.88&11.83 / 32.85 & 14.41 / 38.81&14.13 / 34.32&16.06 / 39.75&17.21 / 37.64&20.34 / 43.40\\\midrule
        (7) EEND-EDA (SSA)&12.95 / 33.98&15.69 / 40.03&12.74 / 34.08&15.86 / 40.44&15.03 / 33.64 & 17.52 / 39.15 & 17.81 / 38.32 & 21.31 / 44.32\\
        (8) TDNN-based x-vector + VBx + EENDasP (SSA)&12.54 / 31.32&14.55 / \textbf{36.11}&12.74 / \textbf{32.20}& 15.34 / \textbf{37.50}&15.45 / 33.61&17.77 / \textbf{38.67}&17.60 / \textbf{35.16} &20.84 / \textbf{40.18}\\
        (9) DOVER-Lap of (1)(2)(4)(7)(8)&\textbf{10.65} / \textbf{30.82}&\textbf{12.47} / 36.21&\textbf{11.58} / 32.37&\textbf{14.09} / 38.25&\textbf{13.85} / \textbf{33.41}&\textbf{15.81} / 38.77&\textbf{16.94} / 36.31&\textbf{20.01} / 41.78\\\bottomrule
    \end{tabular}%
    }
\end{table*}

\autoref{tbl:result} shows the results on the DIHARD III development set and evaluation set.
The results on the evaluation set are from the official scoring server.
Every subsystem significantly outperformed the baseline system \cite{ryant2020third}.
System (5) performed best as a single subsystem without self-supervised adaptation, but the other four subsystems showed the comparable performance.  
Our best system achieved \SI{11.58}{\percent} and \SI{14.09}{\percent} of DERs on the full and core evaluation set in Track 1, respectively.
It also achieved \SI{16.94}{\percent} and \SI{20.01}{\percent} of DERs in Track 2.

\section{Hardware requirements}
\newcommand{\Rmark}{\textsuperscript{\textregistered}\xspace}

We run our experiments using two different infrastructures.
One is equipped with Intel\Rmark Xeon\Rmark CPU Gold 6123 @ \SI{2.60}{\GHz} using up to 56 threads with 750 GB of RAM, and up to eight NVIDIA\Rmark V100\Rmark GPUs with 32 GB of VRAM each and 15.7 single-precision TFLOPS.
Using this infrastructure, we trained and processed the VAD models, the Res2Net models, the PLDA model, the EEND-based systems, and DOVER-Lap.

The other is the JHU's CLSP Cluster, which is equipped with Intel\Rmark Xeon\Rmark CPU E5-2680 v2 @ \SI{2.80}{\GHz} using up to 54 threads and 60GB of RAM, and up to four NVIDIA\Rmark GeForce GTX 1080 Ti\Rmark with 11 GB of VRAM each and 10.6 single-precision TFLOPS.
The TDNN-based extractor, the VBx clustering, and the overlap detection and assignment model were trained on this cluster.

The processing time for WPE dereverberation is \SI{2.54}{\second} for 1 minute of audio.

Our framework's components were trained on PyTorch \cite{adam2019pytorch}, except for the TDNN-based extractor that was trained on Kaldi \cite{povey2011kaldi}.

The SincNet VAD was trained on a single NVIDIA\Rmark V100\Rmark GPU and required about 22 hours for training. The processing of the labels required \SI{0.132}{\second} for 1 minute of audio.

The TDNN VAD was trained with 3 to 8 NVIDIA\Rmark GeForce GTX 1080 Ti\Rmark (We gradually increased the number of GPU jobs during training) for 1 hour.

The TDNN x-vectors, VBx, and the overlap detector extraction were conducted on the CLSP Cluster. The overlap detector required 40 CPUs with a decoding time of 30 mins for all datasets including development and evaluation sets. The TDNN x-vector was trained on 4-8 GPUs and required approximately 48 hours. The PLDAs, trained using CPUs, required around 30 mins to train the VoxCeleb datasets, and 10 mins to train the DIHARD III dataset. The scoring for every file took around \SI{0.25}{\second} for each audio. All the procedures were parallelized using 30 to 40 jobs to reduce the computational time. 

The Res2Net-based x-vector extractors were trained using four NVIDIA\Rmark V100\Rmark GPUs and required 54 hours approximately for training. The processing time for the x-vector extraction using this model is \SI{1.52}{\second} for 1 minute of audio.

The EEND-based models are trained using a single NVIDIA\Rmark V100\Rmark GPU. For EEND-EDA, it took 30 hours for training on Sim2spk, 325 hours for finetuning on the concatenation of Sim1spk to Sim5spk, and 1.5 hours for adaptation on the DIHARD III development set. The processing time of iterative inference and VAD post-processing was about 30 minutes. It takes about 3 hours for self-supervised adaptation, which was almost doubled from the adaptation on the development set because we additionally used the evaluation set with pseudo labels.
For SC-EEND, it took 200 hours for training on simulated mixtures, 2 hours for adaptation, and 5 minutes for inference.

The processing time of EENDasP given the results of TDNN-based x-vectors + VBx was about 5 minutes for the entire development set.

DOVER-Lap of five systems was based on the official repository\footnote{\url{https://github.com/desh2608/dover-lap}}, which took about 3 minutes to process the development set.

The trained models and the generated outputs had a total disk usage of 1.2 TB.
\bibliographystyle{IEEEtran}
\bibliography{ref}
\end{document}